\documentclass[11pt,preprintnumbers,aps,amssymb,nofootinbib,amsmath,superscriptaddress,notitlepage,prd]
{revtex4-2}
\usepackage{epsfig,epsf}
\usepackage{comment}
\usepackage{bm} % puts greek and math symbols in boldface using \bm
\usepackage{color} % {\color{red} ... }
\usepackage{slashed} % Dirac slash
\usepackage{relsize}	%rescalable math 
\usepackage{soul} %spac­ing out (let­terspac­ing), un­der­lin­ing, strik­ing out, etc.
\usepackage{hyperref}
\usepackage{tensor} %\indices{^a_b} produces the properly spaced the tensor indices 
\usepackage{yfonts} %Gotic fonts \textgoth{This: is: Gothic}
\newcommand{\beq}{\begin{equation}}
\newcommand{\beql}[1]{\begin{equation}\label{#1}}
\newcommand{\eeq}{\end{equation}}
\def\bal#1\gal{\begin{align}#1\end{align}}
\newcommand{\ball}[1]{\bal\label{#1}}
\newcommand{\eq}[1]{(\ref{#1})}
\newcommand{\fig}[1]{Fig.~\ref{#1}}
\renewcommand{\sec}[1]{Sec.~\ref{#1}}
\DeclareMathOperator{\Tr}{\mathrm{Tr}}

\DeclareMathOperator{\step}{\mathrm{H}}

\renewcommand{\b}[1]{{\bm #1}} 
\newcommand{\unit}[1]{\hat {{\bm #1}}} % unit vector

\usepackage{feynmp-auto}

\begin{document}

%\title{multiple color chiral radiation}

%\title{Chiral Cherenkov cascades in QED and QCD}

%\title{Time-evolution of parity-odd cascades and jets initiated by a fast particle in homogeneous chiral matter}

\title{Time-evolution of parity-odd cascades in homogeneous Abelian and non-Abelian media with chiral imbalance}

\author{Jeremy Hansen}

\author{Kirill Tuchin}

\affiliation{
Department of Physics and Astronomy, Iowa State University, Ames, Iowa, 50011, USA}

\date{\today}

\begin{abstract}

We study radiation by a fast particle in a medium with a finite chiral chemical potential. The medium’s anomalous response encompasses the chiral magnetic effect, enabling novel chiral Cherenkov and pair production processes. The kinematics of the corresponding cascades is fundamentally different from conventional cascades. Notably, it is not dominated by the strong ordering of momenta. Employing the effective Chern-Simons extensions of QED and QCD to incorporate the chiral magnetic effect, we derive and solve the evolution equations describing the cascades in a chiral medium, presuming that the anomalous contributions are dominant.

\end{abstract}

\maketitle

%%%%%%%%%%%%%%%%%%%%%%%%%%%%%%%%%%%%%%%%
%%%%%%%%%%%%%%%%%%%%%%%%%%%%%%%%%%%%%%%%
\section{Introduction}\label{sec:int}

\begin{figure}[ht]
      \includegraphics[width=0.5\linewidth]{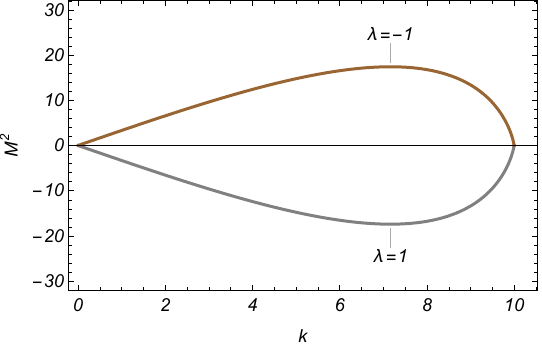} 
  \caption{The squared effective photon mass $M^2=\omega^2-\b k^2$ as a function of photon momentum $|\b k|$ in QED plasma at finite chiral chemical potential $\mu_5$ for different polarizations $\lambda$. We used the dispersion relation derived in \cite{Akamatsu:2013pjd} with $\omega=\mu_5=100\, m_D$. Arbitrary units. }
\label{fig:dispersion}
\end{figure}

Interaction of a fast particle with the medium induces a cascade. Such cascades are studied in many areas of particle and nuclear physics \cite{Arnold:2002zm,Baier:2000sb,Blaizot:2015lma,Mehtar-Tani:2022zwf,Mueller:1999pi,Schlichting:2020lef,PhysRevLett.105.195005,Bulanov:2013cga,Barata:2021wuf,Caron-Huot:2010qjx,Sievert:2019cwq}. This paper studies cascades generated in a medium with chiral fermions. The chiral media are characterized by the anomalous response to the electromagnetic field which gives rise to the chiral magnetic \cite{Kharzeev:2004ey,Kharzeev:2007jp,Kharzeev:2009fn,Kharzeev:2007tn,Fukushima:2008xe} and anomalous Hall effects \cite{Klinkhamer:2004hg,Zyuzin:2012tv,Grushin:2012mt,Kharzeev:2007tn,Kharzeev:2013ffa}.  The anomalous response lifts the chiral degeneracy of the gauge boson dispersion relation, which acquires dependence on the gauge boson polarization. In particular, it exhibits two distinct branches: one is spacelike and the other is timelike as seen in \fig{fig:dispersion}. As a result the  $1\to 2$ and $2\to 1$ processes,  prohibited in the free space by the energy-momentum conservation, become  allowed in the chiral medium. 

The dynamics of the gauge theories at finite chiral chemical potential $\mu_5$ can be conveniently described by adding the Chern-Simons term $-\theta F\tilde F/4$ to the the Lagrangian \cite{Wilczek:1987mv,Sikivie:1984yz} and requiring that the time derivative of the pseudo-scalar field be constant: $\partial_0\theta=\mu_5=b_0/c_A$, where $c_A$ is the  anomaly coefficient and $b_0$ is dubbed the chiral magnetic conductivity \cite{Kharzeev:2009pj,Fukushima:2008xe}, and that its spacial gradients vanish: $\b\nabla\theta=0$. The latter equation reflects the assumed uniformity of the chiral medium. We will refer to such an extension of QED and QCD as $\chi$QED and $\chi$QCD respectively. Unlike the original theories, the $\chi$-extensions are parity-odd. As a result, the rates of processes involving the gauge bosons depend on their polarizations. The Chern-Simons term causes emergence of the effective photon mass as seen in \fig{fig:dispersion}. In particular, at photon momenta much larger than the plasma frequency the dispersion relation takes form 
\ball{i0}
\omega^2= \b k^2 -\lambda b_0 |\b k|\,,
\gal
where $k=(\omega,\b k)$ and $\lambda=\pm 1$ indicate  the right or left-hand polarization respectively.

In the dilute regime, only $1\to 2$ decays play essential role in the cascade. The decay $e\to e+\gamma$ in $\chi$QED is known as the chiral Cherenkov radiation \cite{Tuchin:2018sqe,Huang:2018hgk,Tuchin:2018mte,Barredo-Alamilla:2023xdt}\footnote{ It is closely related to the vacuum Cherenkov radiation \cite{Kostelecky:2002ue,Lehnert:2004hq,Lehnert:2004be}.} because like the conventional Cherenkov radiation it has infinite coherence length. Only one---right or left-handed---circular photon polarization can be produced by the chiral Cherenkov radiation, namely, the polarization corresponding to the spacelike dispersion relation. On the other hand, the pair-production decay $\gamma\to e+\bar e$ proceeds only if the photon is timelike which corresponds to the opposite polarization. As a result, an electromagnetic cascade promptly develops 100\% chirality. We will refer to all decays induced by the chiral anomaly collectively as \emph{the chiral Cherenkov decays} and to the corresponding cascade as \emph{the chiral Cherenkov cascade}.

The dynamics of the chiral Cherenkov cascade in $\chi$QCD is more complicated because along with the quasi-Abelian processes $q\to q+g$ and $g\to q+\bar q$ it allows the genuine non-Abelian decay $g\to g+g$, which necessarily involves gluons of different polarizations \cite{Hansen:2024rlj}. 

The mathematical description of the conventional cascades in QED and QCD significantly simplifies at high energies because the individual decays factor out thanks to the strongly ordered transverse momenta. This makes the decays quasi-classical, and thereby statistically independent at the leading logarithmic approximation. Unlike the conventional decays, the chiral Cherennkov decays are not strongly ordered in the transverse momentum. Rather the transverse momenta of the radiated particles are fixed by the kinematic constraints. Nevertheless, precisely this fixing of the transverse momenta makes the factorization and the associated quasi-classical interpretation possible for the chiral Cherenkov cascade as we argue in \sec{sec:dbl}. 
%\footnote{The instability of the infrared modes $k<b_0$ of the electromagnetic field  can be ignored at high energies \cite{Tuchin:2018sqe}.}

Once the factorization property is established, the dynamics of the chiral Cherenkov cascade can be described by the kinetic equation. Consider the chiral Cherenkov cascade in $\chi$QED. Introduce the electron, positron, right-handed photon and left-handed photon distributions $e(x,t)$, $\bar e(x,t)$, $\gamma^+(x,t)$ and $\gamma^-(x,t)$ respectively, where $x$ denotes the longitudinal momentum fraction carried by the decay product and $t$ is time. In realistic systems the chiral Cherenkov processes compete with the conventional ones. However, in this paper, we will ignore the conventional processes in order to obtain the most clear physical picture of the chiral Cherenkov cascade. Throughout the paper we assume for definitiveness that $b_0>0$; as a result the decay $e\to e+\gamma^+$ produces the right-handed photons, whereas the pair-production $\gamma^-\to e+\bar e$  destroys the left-handed ones. We note that the distributions $e$ and $\bar e$ obey the same kinetic equation, although possibly with different initial conditions, given that the kinematic constraints due to the anomaly are unaffected by the sign of electron's electric charge. Therefore, the general structure of the kinetic equations for the chiral Cherenkov cascade in $\chi$QED is
\begin{subequations}\label{i1}
    \bal    
     e(y,t+\Delta t)&= \left(1-W_\text{tot}^{e\to e\gamma^+}\Delta t\right)e(y,t)+ \Delta t \int_0^1 dz\int_0^1 dx  \frac{dW^{e\to e\gamma^+}}{dx}e(z,t)\delta(y-xz)\nonumber\\
       &+
       \Delta t \int_0^1 dz\int_0^1 dx  \frac{dW^{\gamma^-\to e\bar e}}{dx}\gamma^-(z,t) \delta(y-xz)\label{i2} \\
       \bar e(y,t+\Delta t)&= \left(1-W_\text{tot}^{\bar e\to \bar e\gamma^+}\Delta t\right)\bar e(y,t)+ \Delta t \int_0^1 dz\int_0^1 dx  \frac{dW^{\bar e\to \bar e\gamma^+}}{dx}\bar e(z,t)\delta(y-xz)\nonumber\\
       &+
       \Delta t \int_0^1 dz\int_0^1 dx  \frac{dW^{\gamma^-\to e\bar e}}{dx}\gamma^-(z,t) \delta(y-xz)\,,\label{i3}
\\     
      \gamma^+(y,t+\Delta t)&= \gamma^+(y,t)+\Delta t \int_0^1 dz\int_0^1 dx  \frac{dW^{e\to \gamma^+ e}}{dx}\left[ e(z,t)+\bar e(z,t)\right] \delta(y-xz)\,,\label{i4}\\
       \gamma^-(y,t+\Delta t)&= \left( 1- W_\text{tot}^{\gamma^-\to e\bar e}\Delta t \right) \gamma^+(y,t)\,,\label{i5}
    \gal
\end{subequations}
where $dW^{a\to bc}/dx$ is the inclusive rate of the indicated process and $W^{a\to bc}_\text{tot}$ is the corresponding total rate. Unlike the inclusive rate, the total one includes the virtual contributions which are essential to maintain the charge and energy conservation and the gauge invariance. The terms in the right-hand side of \eq{i2} have the following meaning: the expression in the parentheses in the first term is the probability that no photon is radiated during the time interval $\Delta t$, the second term describes the probability that the parent electron emits a photons and ends up with the energy fraction $y$, and the third term is the pair production. Since the right-handed photon cannot decay, there is no term proportionate to $\gamma^+$ in the right-hand side of \eq{i4}. In contrast, in \eq{i5} there are no gain terms because the left-handed photons can only decay, but not being produced. The differential equations are obtained by taking the limit $\Delta t\to 0$. The kinetic equations in $\chi$QCD can be obtain in a similar way. The derivation and solution of these equations is the main subject of this paper.

The paper is organized as follows. Secs.~\ref{sec:spe}--\ref{sec:kinetic} deal with the chiral Cherenkov cascade in QED and in the remaining sections -- in QCD. In \sec{sec:spe} we quote the inclusive single particle rates for $\chi$QED followed in \sec{sec:dbl} by the analysis of the double inclusive production where we prove the factorization of the cascade into the product of the single inclusive processes. In \sec{sec:kinetic} we derive the kinetic equations for $\chi$QED---assuming that the chiral Cherenkov decays are dominant---and solve them numerically for special choices of the initial conditions. In \sec{sec:QCD} we derive and solve the kinetic equations of $\chi$QCD. The summary, conclusions and outlook are presented in \sec{sec:summary} where we also discuss how the chiral Cherenkov decays can be included into the existing models of cascades and jets.

%%%%%%%%%%%%%%%%%%%%%%%%%%%%%%%%%%%%%%%%
%%%%%%%%%%%%%%%%%%%%%%%%%%%%%%%%%%%%%%%%
\section{Single parton emission in $\chi$QED}\label{sec:spe}

One of the two basic processes of the chiral Cherenkov cascade in $\chi$QED is the photon production $e\to e\gamma$.  Assume for  simplicity that the plasma frequency is negligible $\omega_p=0$ and denote $p=(E,\b p)$, $p'=(E',\b p')$, $k=(\omega,\b k)$ the momenta of the initial and final fermion and photon respectively. The rate of the right-handed photon production is \cite{Tuchin:2018sqe}:
\bal
\frac{dW^{e\to \gamma^+ e}}{dx}&= \frac{\alpha b_0}{4}\left( \frac{(1-x)^2+1}{x}-\frac{2m^2}{b_0E}\right)\step(x\le x_0)\,,\label{p1}
\gal
where $x=\omega/E$ is fraction of the incident energy carried away by the photon,
\bal\label{p2}
x_0=\frac{1}{1+\frac{m^2}{b_0E}}\,,
\gal
and the step function $\step(M)$ equals unity if the condition $M$ is satisfied and vanishes otherwise. Let now $x$ denote the energy fraction carried away by the final electron, i.e.\ $x=E'/E$. Then the electron production rate reads 
\bal
\frac{dW^{e\to e\gamma^+}}{dx}&= \frac{\alpha b_0}{4}\left( \frac{x^2+1}{1-x}-\frac{2m^2}{b_0E}\right)\step(x\ge 1-x_0)\,.\label{p4}
\gal

Another basic process is the pair production $\gamma^-\to e\bar e$. Let $k$, $p$ and $p'$ now refer to the momenta of the initial photon and final electron and positron respectively and  $x=E/\omega$ be the fraction of the incident photon energy carried away by electron. The electron production rate is given by \cite{Tuchin:2018sqe}
\bal
\frac{dW^{\gamma^-\to e\bar e}}{dx}&= \frac{\alpha b_0}{4}\left( x^2+(1-x)^2+\frac{2m^2}{\omega b_0}\right)\step(x_1\le x\le x_2) \,,\label{p6}
\gal
where 
\bal\label{p7}
x_{1,2}= \frac{1}{2}\left(1\mp \sqrt{1-\frac{4m^2}{b_0\omega}}\right)\,.
\gal

It is convenient to introduce the following parameters: the characteristic time scale $\tau$ such that
\bal\label{p12}
\frac{1}{\tau}= \frac{\alpha b_0}{4}\,,
\gal
and the dimensionless mass parameter 
\bal\label{p14}
\nu=\frac{m^2}{b_0E_0}\,,
\gal
where $E_0$ is the total cascade energy.

%%%%%%%%%%%%%%%%%%%%%%%%%%%%%%%%%%%%%%%%%%%%%%%%%%%%%%%%%%%%%%%%%%%%%%%
\section{Double particle emission in $\chi$QED}\label{sec:dbl}

We are now going to argue that the multiparticle production rate can be represented as a convolution of the single particle production rates. We first consider the double photon emission as a specific example.   

The general expression for the single photon emission rate reads:
\ball{k0}
W^{e\to e\gamma}= \frac{1}{2E}\frac{1}{2}\sum_\text{spins}\sum_{\lambda}
\int\frac{d^3p'}{2E'(2\pi)^3}\int\frac{d^3k}{2\omega(2\pi)^3}
(2\pi)^4\delta(p-k-p')
\left|e\bar u'\slashed{\epsilon}^* u\right|^2\,.
\gal
We use the shorthand notation $u\equiv u(\b p,s)$, $u'\equiv u(\b p',s')$, $\epsilon\equiv \epsilon(\b k, \lambda)$, with the bold letters referring to the spacial components of the corresponding four-vectors, $s$ and $s'$ label fermion spin states, and $\lambda$ photon polarizations. 

\begin{figure}[ht]
      \includegraphics[width=12cm]{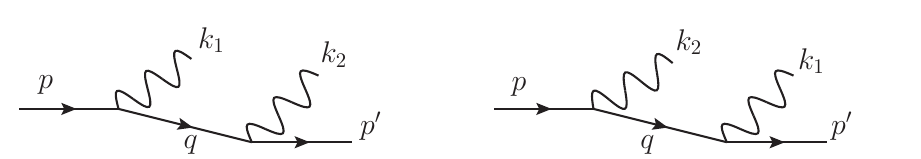} 
  \caption{The diagrams for double photon production.}
\label{fig:dbl-photon}
\end{figure}

The rate for the two photon emissions by an electron reads 
\bal\label{k1}
W^{e\to e\gamma\gamma}= 
\frac{1}{8E}\sum_\text{spins}\sum_{\lambda_1,\lambda_2}\int\frac{d^3p'}{2E'(2\pi)^3}\int\frac{d^3k_1}{2\omega_1(2\pi)^3}\int\frac{d^3k_2}{2\omega_2(2\pi)^3}(2\pi)^4\delta(p-k_1-k_2-p')
\nonumber\\
\times\left|\frac{e^2\bar u'\slashed{\epsilon}^*_2 (\slashed p-\slashed k_1+m) \slashed{\epsilon}^*_1 u}{(p-k_1)^2-m^2+2iE_q/\tau}
+
\frac{e^2\bar u'\slashed{\epsilon}^*_1 (\slashed p-\slashed k_2+m) \slashed{\epsilon}^*_2 u}{(p-k_2)^2-m^2+2iE_q/\tau}
\right|^2\,,
\gal
where  $E_q= E-\omega_1$, $\tau$ is the relaxation time controlling the resonance width \cite{Hansen:2022nbs,Hansen:2023wzp} and the symmetry factor is $1/2$. In the ultrarelativistic approximation, the denominator of the propagator in the first term can be written as \cite{Hansen:2023wzp}
\ball{k3}
\frac{1}{-2p\cdot k_1+k_1^2+2iE_q/\tau}= \frac{1}{\omega_1 E\left( \frac{m^2}{E^2\omega_1}\frac{\omega_1-x_0 E}{1-x_0}+\vartheta^2\right)+2iE_q/\tau}\,.
\gal
Apparently, when $\omega_1<x_0 E$ the propagator has a resonance at a finite photon emission angle $\vartheta$. The second term in \eq{k1} has  a similar resonance. Since the resonance contribution dominates the rate we can approximate 
\ball{k5}
\left|\frac{1}{(p-k_1)^2-m^2+2iE_q/\tau}\right|^2\approx \frac{\pi\tau}{2E_q} \delta\left((p-k_1)^2-m^2\right)\,.
\gal
In other words, the intermediate fermion is effectively on the mass-shell. A more detailed justification of the approximation \eq{k5} is presented in Appendix~\ref{Adpe}. We argue there that the relaxation time must satisfy $\tau\gg 1/(2\pi b_0)$. 
Eq.~\eq{k5} allows us to write the rate of the double photon production as the product of two single photon rates. Writing 
$$
\delta(p-k_1-k_2-p')= \int d^4q \delta(q-p+k_2)\delta(q-p'-k_2)
$$
and $\slashed p-\slashed k_1+m= \sum_s u(\b q,s)\bar u(\b q,s)$ we obtain employing \eq{k0}
\bal\label{t7}
W^{e\to e\gamma\gamma}= \tau \int_0^E d\omega_1 \frac{dW^{e(p)\to e(q)\gamma(k_1)}}{d\omega_1}\int_0^{E_q}d\omega_2 \frac{dW^{e(q)\to e(p')\gamma(k_2)}}{d\omega_2}\,.
\gal
 The other branching process --- the pair production --- factors out in the same way. 

This derivation can be readily generalized to an arbitrary multiparticle process. The systematic way to describe such a cascade is to employ the evolution equations which we develop in the following sections.

%%%%%%%%%%%%%%%%%%%%%%%%%%%%%%%%%%%%%%%%%%%%%%%%%%%%%%%%%%%%%%%%%%%%%%%%%%%%%%%%%%%%
%%%%%%%%%%%%%%%%%%%%%%%%%%%%%%%%%%%%%%%%%%
\section{Chiral Cherenkov cascade in $\chi$QED}\label{sec:kinetic}

The distribution functions of electrons, positrons and photons are $e(x)$, $\bar e(x)$, and $\gamma^\pm(x)$  respectively. They indicate the probability to find a particle with the energy fraction $x$ of the total cascade energy and with any transverse momentum.  We assume that the particle densities are small so that the linear approximation applies. We also neglect all conventional contributions to the cascades. 

\subsection{Evolution equations in chiral limit $m=0$.}

According to \eq{p1} and \eq{p6}, in the chiral limit, the rates of various production processes are 
\begin{subequations}\label{N1}
\bal
\frac{dW^{e\to \gamma^+ e}}{dx}&= \frac{1}{\tau}\frac{1+(1-x)^2}{x}\,,\label{n1}\\
\frac{dW^{e\to e\gamma^+}}{dx}&= \frac{1}{\tau}\frac{1+x^2}{1-x}\,,\label{n2}\\
\frac{dW^{\gamma^-\to e\bar e}}{dx}&= \frac{1}{\tau}[x^2+(1-x)^2]\,,\label{n3}
\gal
\end{subequations}
where $x$ is the fraction of energy carried by the outgoing particle listed first. The corresponding splitting functions including the virtual terms are
\begin{subequations}\label{N8}
\bal
P_{\gamma^+ e}(x)&= \frac{1+(1-x)^2}{x}\,,\label{n8}\\
P_{e\gamma^-}(x)&= x^2+(1-x)^2\,,\label{n9}\\
P_{ee}(x)&= \frac{1+x^2}{(1-x)_+}+\frac{3}{2}\delta(1-x)\,,\label{n10}\\
P_{\gamma^-\gamma^-}&= -\frac{2}{3}\delta(1-x)\,,\label{n11}
\gal
\end{subequations}
The distribution $1/(1-x)_+$ in \eq{n10} is defined as usual: 
\bal\label{n13}
\int_x^1dz\frac{f(z)}{(1-z)_+}=
\int_x^1dz\frac{[f(z)-f(1)]}{1-z}+f(1)\ln(1-x)\,.
\gal

The evolution equations \eq{i1} now read:
\begin{subequations}\label{N20}
\bal
\frac{de(y,t)}{dt}=&\frac{1}{\tau} \int_y^1\frac{dx}{x}\left\{ P_{ee}(x)e\left( \frac{y}{x},t\right)+P_{e\gamma^-}(x)\gamma^-\left( \frac{y}{x},t\right)\right\}\,,\label{n20}\\
\frac{d\bar e(y,t)}{dt}=&\frac{1}{\tau} \int_y^1\frac{dx}{x}\left\{ P_{ee}(x)\bar e\left( \frac{y}{x},t\right)+P_{e\gamma^-}(x)\gamma^-\left( \frac{y}{x},t\right)\right\}\,,\label{n21}\\
\frac{d\gamma^+(y,t)}{dt}=&\frac{1}{\tau} \int_y^1\frac{dx}{x}\left\{ P_{\gamma^+ e}(x)e\left( \frac{y}{x},t\right)+P_{\gamma^+ e}(x)\bar e\left( \frac{y}{x},t\right)\right\}\,,\label{n22}\\
\frac{d\gamma^-(y,t)}{dt}=&\frac{1}{\tau} \int_y^1\frac{dx}{x} P_{\gamma^- \gamma^-}(x)\gamma^-\left( \frac{y}{x},t\right)\,.\label{n23}
\gal
\end{subequations}
where $\tau$ is given by \eq{p12}. These equations are similar to the evolution equations of QED. The main differences are (i) Eqs.~\eq{N20} explicitly depend on the photon polarization, i.e.\ $\gamma^+$ and $\gamma^-$ evolve differently, and (ii) $t$ is a linear rather than a logarithmic scale. 

One can write these equations in terms of the moments (Mellin transform)
\bal\label{n25}
f_n(t)= \int_0^1 x^{n-1} f(x,t) dx\,,
\gal
as follows:
\begin{subequations}\label{N27}
\bal
\frac{de_n(t)}{dt}= &\frac{1}{\tau}\left\{ A_n^{ee}e_n(t)+ A_n^{e\gamma^-}\gamma^-_n(t)\right\}\,,\label{n27}\\
\frac{d\bar e_n(t)}{dt}= &\frac{1}{\tau}\left\{ A_n^{ee}\bar e_n(t)+ A_n^{e\gamma^-}\gamma^-_n(t)\right\}\,,\label{n28}\\
\frac{d\gamma^+_n(t)}{dt}= &\frac{1}{\tau}\left\{ A_n^{\gamma^+ e}e_n(t)+ A_n^{\gamma^+ e}\bar e_n(t)\right\}\,,\label{n29}\\
\frac{d\gamma^-_n(t)}{dt}= &\frac{1}{\tau} A_n^{\gamma^- \gamma^-}\gamma^-_n(t)\,,\label{n30}
\gal
\end{subequations}
where 
\bal\label{n32}
 A_n^{ab}= \int_0^1 x^{n-1} P_{ab}(x)dx\,.
\gal
In this form it is easy to check conservation of the electric charge and energy of the cascade. The electric charge in units of the electron charge is given by 
\bal\label{n34}
\int_0^1 \left\{ e(x,t)-\bar e(x,t)\right\} dx= e_1(t)-\bar e_1(t)\,.
\gal
Using \eq{n27} and \eq{n28} we obtain
\bal
\frac{dQ}{dt}= \frac{1}{\tau}A_1^{ee}\left\{ e_1(t)-\bar e_1(t)\right\}=0 \label{n35}
\gal
because 
\bal
A_1^{ee}= \int_0^1  P_{ee}(x) dx=0\,. \label{n36}
\gal
Similarly the total energy of the cascade is 
\bal
\int_0^1 x\left\{ e(x,t)+\bar e(x,t)+\gamma^+(x,t)+ \gamma^-(x,t)\right\} dx=e_2(t)+ \bar e_2(t) + \gamma^+_2(t) + \gamma_2^-(t)\,. \label{n38}
\gal
That the time derivative of the total energy vanishes is derived by substituting \eq{n27}--\eq{n30} and noting that 
\begin{subequations}
\bal
&A_2^{ee}+ A_2^{\gamma^+ e}= \int_0^1 x\left( P^{ee}(x)+ P^{\gamma^+ e}(x)\right) dx=0\,,\label{n39}\\
& 2A_2^{e\gamma^-}+A_2^{\gamma^-\gamma^-}= \int_0^1 x\left( 2P^{e\gamma^-}(x)+P^{\gamma^-\gamma^-}(x)\right)=0\,. \label{n40}
\gal
\end{subequations}

Let the initial conditions at $t=0$ be 
\ball{m7}
e(x,0)&=\delta(1-x)\,,\quad \bar e(x,0)=\gamma^+(x,0)=\gamma^-(x,0)=0\,,
\gal
The general solution to \eq{n23} is 
\ball{m11}
\gamma^-(x,t)&= \gamma^-(x,0)e^{-2t/3\tau}\,,
\gal 
In view of the initial conditions \eq{m7}, the distribution of the negatively polarized photons vanishes $\gamma^-(x,t)=0$. It follows from \eq{n28} that $\bar e(x,t)=0$. The solution to \eq{n27} is
\bal\label{m13}
e_n(t)= \exp\frac{A_n^{ee} t}{\tau}\,,
\gal
where 
\bal\label{m15}
A_n^{ee}= \frac{3}{2}-\frac{2n+1}{n(n+1)}\,.
\gal
To obtain the electron distribution one performs the inverse Mellin transform:
\bal
e(x,t)= \frac{1}{2\pi i}\int_{-i\infty}^{+i\infty} e_n(t)x^{-n}dn\,,\label{m17}
\gal
where the contour runs parallel to the imaginary axis to the right of the singularities of the integrand. Once $e(x,t)$ is known, it is easy to obtain $\gamma^+(x,t)$ from \eq{n22}. Unfortunately, the integral \eq{m17} cannot be done analytically.

\begin{figure}[ht]
\begin{tabular}{cc}
      \includegraphics[width=0.45\linewidth]{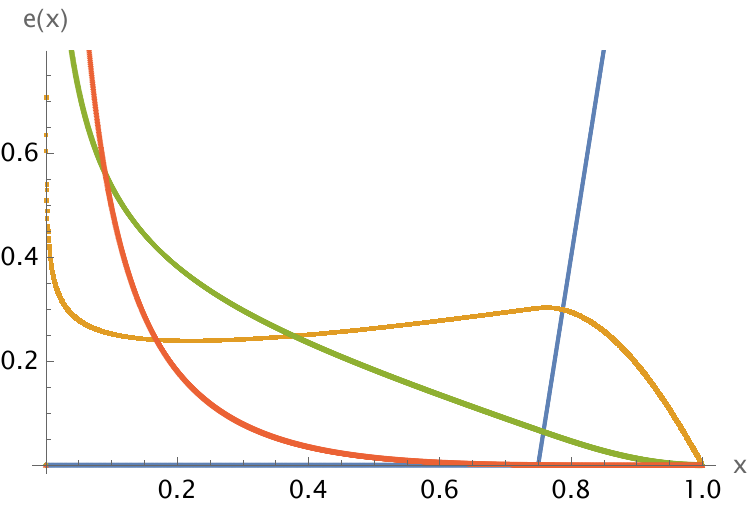} &
      \includegraphics[width=0.45\linewidth]{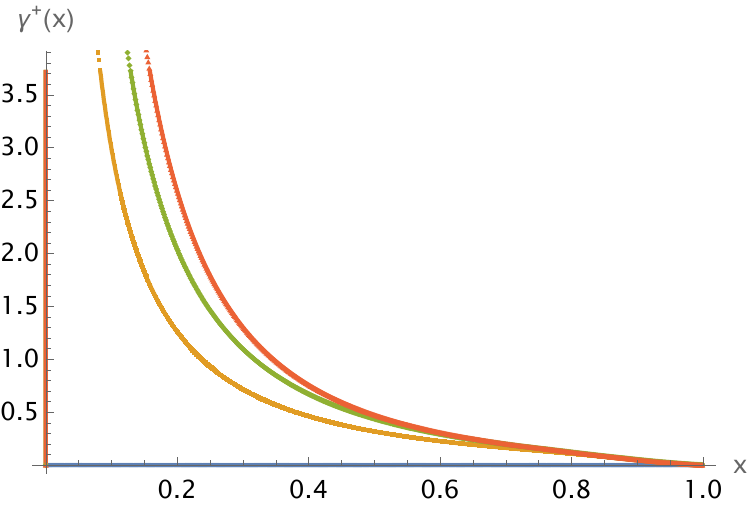}
      \end{tabular}
  \caption{Chiral Cherenkov cascade in $\chi$QED with the initial conditions: $e(x,0)$ given by \eq{m20}, and  $\bar e(x,0)=\gamma^\pm (x,0)=0$. Blue line: $t=0$, yellow line: $t=\tau/2$, green line: $t=\tau$, red line: $t=2\tau$. }
\label{fig:electron}
\end{figure}

The evolution equations can be solved numerically. We  use the procedure developed in \cite{Hirai:1997gb}. We first consider the initial conditions similar to \eq{m7} except that we replace the delta-function by  a smoother distribution:
\ball{m20}
e(x,0)= 32 \left(x-\frac{3}{4}\right)\step\left(x\ge \frac{3}{4}\right)\,,
\gal
normalized such that $\int_0^1 e(x,0)dx=1$. The results are shown in Figs.~\ref{fig:electron}  Once can see rapid increase with time of the soft ($x\ll 1$) right-handed photons and the concurrent shift of the electron distribution towards smaller values of $x$. As expected, the cascade is 100\% polarized. 

\begin{figure}[ht]
\begin{tabular}{cc}
      \includegraphics[width=0.45\linewidth]{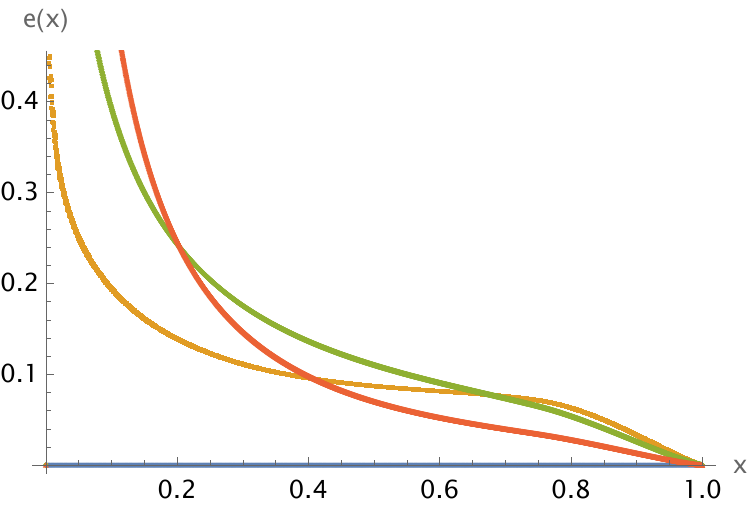} &
      \includegraphics[width=0.45\linewidth]{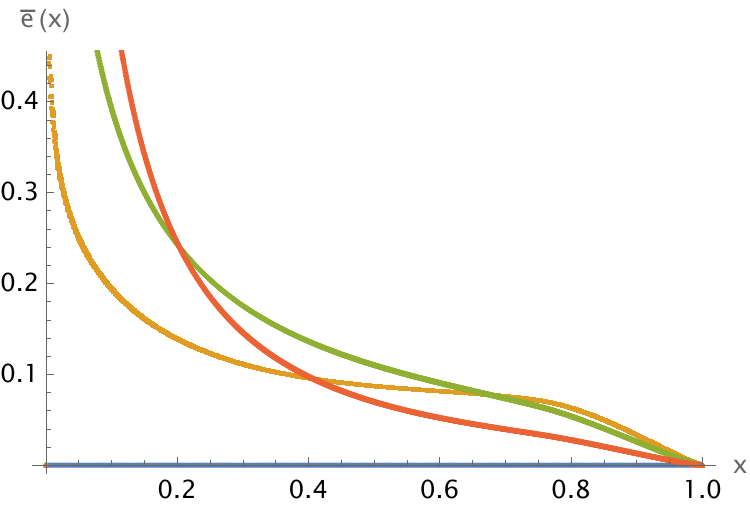}\\
      \includegraphics[width=0.45\linewidth]{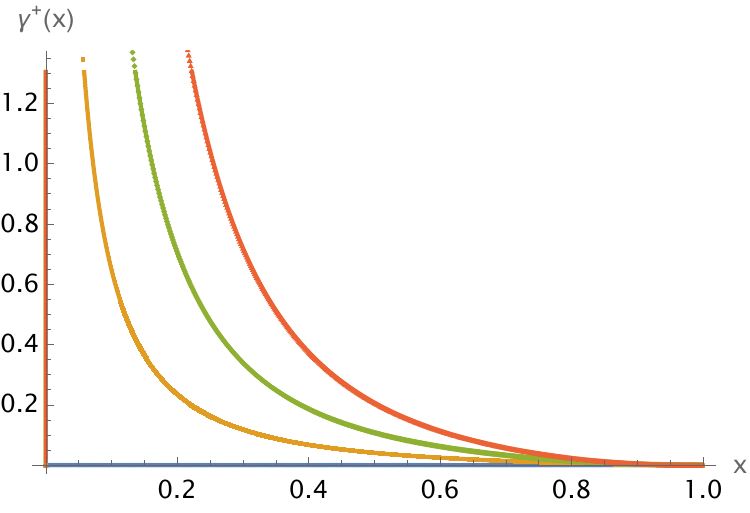} &
      \includegraphics[width=0.45\linewidth]{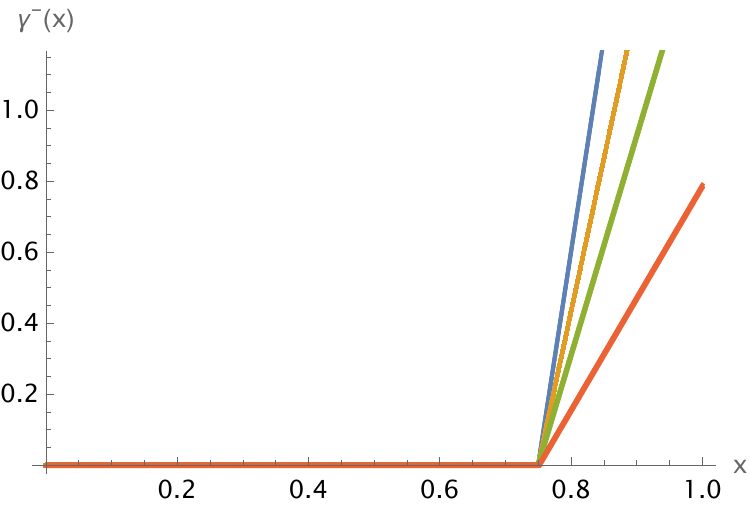}
      \end{tabular}
  \caption{Chiral Cherenkov cascade in $\chi$QED with the initial conditions \eq{m22-S}. Blue line: $t=0$, yellow line: $t=\tau/2$, green line: $t=\tau$, red line: $t=2\tau$. }
\label{fig:neglight}
\end{figure}
\begin{figure}[ht]
\begin{tabular}{cc}
      \includegraphics[width=0.45\linewidth]{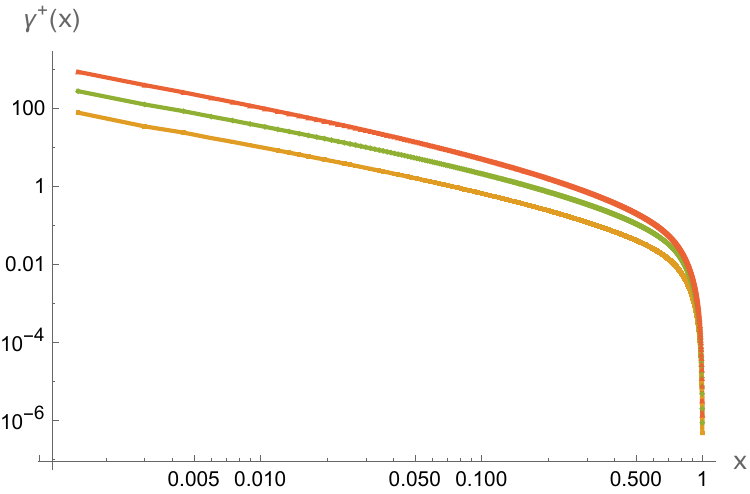} &
      \includegraphics[width=0.45\linewidth]{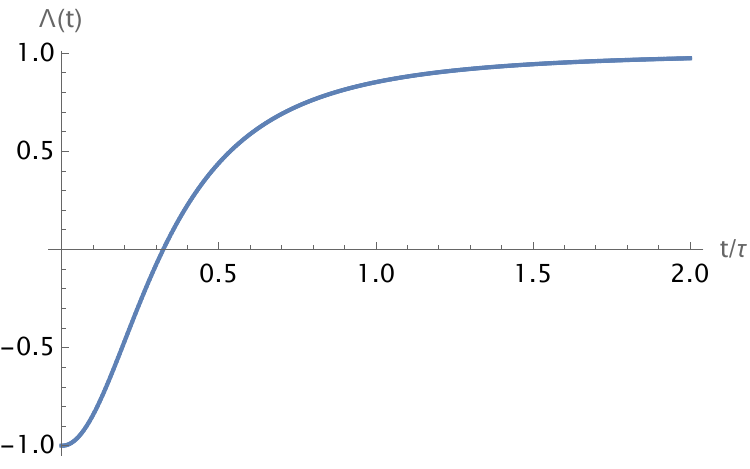}
      \end{tabular}
  \caption{Left panel: log plot for $\gamma^+$. Right panel: The degree of polarization $\Lambda(t)$  of the cascade shown in \fig{fig:neglight}.}
\label{fig:lambda_e}
\end{figure}

\fig{fig:neglight} exhibits evolution of the cascade initiated by the left-handed photons with a different initial distribution 
\begin{subequations}\label{m22-S}
\bal
\gamma^-(x,0)&= 32 \left(x-\frac{3}{4}\right)\step\left(x\ge \frac{3}{4}\right)\,,\label{m22}\\
\gamma^+(x,0)&= e(x,0)= \bar e(x,0)=0\,.\label{m23}
\gal
\end{subequations}
It corresponds to the left-handed photon beam.
The left-handed photons produce the electron-positron pairs which in term radiate the right-handed photons. As a result the number of the right-handed photons increases, while the number of the left-handed ones decreases. The rapid increase of the number of the right-handed photons is especially clearly seen in the left panel of \fig{fig:lambda_e}. 

The chirality distribution of the cascade is proportional to $\gamma^+(x,t)-\gamma^-(x,t)$. 
To characterize the cascade polarization, define \emph{the degree of the cascade polarization}:    
\ball{m25}
\Lambda(t) = \frac{\int_0^1 \left(\gamma^+(x,t)-\gamma^-(x,t) \right)dx}{ \int_0^1 \left(\gamma^+(x,t)+\gamma^-(x,t) \right)dx}\,.
\gal
For the cascade with the initial condition \eq{m22-S} it is displayed in the right panel of \fig{fig:lambda_e} which indicates gradual polarization of the cascade. The cascade starts as completely left-handed and by the time $t\sim \tau$ it is nearly entirely right-handed.

%%%%%%%%%%%%%%%%%%%%%%%%%%%%%%
\subsection{Evolution equations at finite $m$.}

At finite electron mass $m$ the rates \eq{p1} and \eq{p6} read
\begin{subequations}\label{o1A}
\bal
\frac{dW^{e\to \gamma^+ e}}{dx}&= \frac{1}{\tau}\Pi_{\gamma^+ e}\left(x,\frac{y}{x}\right)\,,\label{o1}\\
\frac{dW^{e\to e\gamma^+}}{dx}&= \frac{1}{\tau}\Pi_{ee}\left(x,\frac{y}{x}\right)\,,\label{o2}\\
\frac{dW^{\gamma^-\to e\bar e}}{dx}&= \frac{1}{\tau}\Pi_{e\gamma^-}\left(x,\frac{y}{x}\right)\,.\label{o3}
\gal
\end{subequations}
We defined the following generalizations of the splitting functions:
\begin{subequations}\label{o6A}
\bal
\Pi_{\gamma^+ e}\left(x,z\right)&= \left[ P_{\gamma^+ e}(x)-\frac{2\nu }{z}\right] \step\left(x\le x_0\left(z\right)\right) \,,\label{o6}\\
\Pi_{ee}\left(x,z\right)&=\left[ \frac{1+x^2}{(1-x)_+}+C_e\left(z\right)\delta(1-x)-\frac{2\nu }{z}\right]\step\left(x\ge 1-x_0\left(z\right)\right)\,,\label{o7}\\
\Pi_{e\gamma^-}\left(x,z\right)&=\left[ P_{e\gamma^-}(x)+\frac{2\nu }{z}\right] \step\left(x_1\left(z\right)\le x\le x_2\left(z\right)\right)\,,\label{o8}\\
\Pi_{\gamma^-\gamma^-}(x,z)&= C_\gamma (z)\delta(1-x)\,,\label{o9}
\gal
\end{subequations}
where $\nu$ is given by \eq{p14}, $C_e(z)$, $C_\gamma(z)$ are given by \eq{o38A}, the step function $\step(M)$ equals unity if the condition $M$ is satisfied and vanishes otherwise, and
\bal
x_0\left(z\right)&= \frac{1}{1+\frac{\nu }{z}}\,,\label{o10}\\
x_{1,2}\left(z\right)&= \frac{1}{2}\left( 1\mp\sqrt{1-\frac{4\nu}{z}}\right)\,.\label{o11}
\gal

The evolution equations read
\begin{subequations}\label{o20A}
\bal
\frac{de(y,t)}{dt}=&\frac{1}{\tau} \int_y^1\frac{dx}{x}\left\{ \Pi_{ee}\left(x,\frac{y}{x}\right)e\left( \frac{y}{x},t\right)+\Pi_{e\gamma^-}\left(x,\frac{y}{x}\right)\gamma^-\left( \frac{y}{x},t\right)\right\}\,,\label{o20}\\
\frac{d\bar e(y,t)}{dt}=&\frac{1}{\tau} \int_y^1\frac{dx}{x}\left\{ \Pi_{ee}\left(x,\frac{y}{x}\right)\bar e\left( \frac{y}{x},t\right)+\Pi_{e\gamma^-}\left(x,\frac{y}{x}\right)\gamma^-\left( \frac{y}{x},t\right)\right\}\,,\label{o21}\\
\frac{d\gamma^+(y,t)}{dt}=&\frac{1}{\tau} \int_y^1\frac{dx}{x}\left\{ \Pi_{\gamma^+ e}\left(x,\frac{y}{x}\right)e\left( \frac{y}{x},t\right)+\Pi_{\gamma^+ e}\left(x,\frac{y}{x}\right)\bar e\left( \frac{y}{x},t\right)\right\}\,,\label{o22}\\
\frac{d\gamma^-(y,t)}{dt}=&\frac{1}{\tau} \int_y^1\frac{dx}{x}\, \Pi_{\gamma^- \gamma^-}\left(x,\frac{y}{x}\right)\gamma^-\left( \frac{y}{x},t\right)\,.\label{o23}
\gal
\end{subequations}
The functions $\Pi_{ab}$ reduce to the  splitting functions $P_{ab}$ in the chiral limit $\nu\to 0$.

The coefficients $C_e$, $C_\gamma$ are determined from the charge and energy conservation. To fix them we proceed along the same lines as in the chiral limit. Multiplying the difference of \eq{o20} and \eq{o21} by $y^{n-1}$ and integrating over $0\le y\le 1$ we obtain:
\bal\label{o25}
\frac{d}{dt}\int_0^1 dy y^{n-1}\left\{ e(y,t)-\bar e(y,t)\right\}
=\int_0^1 dy y^{n-1}\int_y^1\frac{dx}{x}\Pi_{ee}\left(x,\frac{y}{x}\right)\left\{ e\left( \frac{y}{x},t\right)-\bar e\left( \frac{y}{x},t\right)\right\}
\gal
Changing the order of integrations $\int_0^1 dy \int_y^1 dx= \int_0^1 dx\int _0^x dy$ and changing integration variables to $x $  and $z=y/x$ we get
\bal\label{o27}
\frac{d}{dt}\int_0^1 dy y^{n-1}\left\{ e(y,t)-\bar e(y,t)\right\}
=\int_0^1 dz  z^{n-1}\mathcal{A}_n^{ee}(z)\left\{ e\left( z,t\right)-\bar e\left(z,t\right)\right\}\,,
\gal
where 
\bal\label{o28}
\mathcal{A}_n^{ee}(z)= \int_0^1 dx x^{n-1} \Pi_{ee}\left(x,z\right)\,.
\gal
When $\nu\to 0$, $\mathcal{A}_n^{ee}(z)\to A_n^{ee}$. However, at finite $\nu$, $\mathcal{A}_n^{ee}(z)$ explicitly depends on $z$ thus breaking the factorization in the Mellin space. In particular, the evolution equations for the moments do not reduce to the set of  ordinary differential equations as was in the chiral case \eq{N27}. Nevertheless, Eq.~\eq{o27} with $n=1$ indicates that $\dot Q=0$ if
\ball{o30}
\mathcal{A}_1^{ee}(z)= 0\,
\gal
for any $z$. The converse statement is not true, viz.\ \eq{o30} does not necessarily follow from the charge conservation,  since the expression in the curly brackets in \eq{o27} is not positive-definite. Similarly, one can prove that the total energy is conserved iff 
\begin{subequations}
\bal
&\mathcal{A}_2^{ee}(z)+ \mathcal{A}_2^{\gamma^+ e}(z)= \int_0^1 x\left( \Pi^{ee}(x,z)+ \Pi^{\gamma^+ e}(x,z)\right) dx=0\,,\label{o33}\\
& 2\mathcal{A}_2^{e\gamma^-}(z)+\mathcal{A}_2^{\gamma^-\gamma^-}(z)= \int_0^1 x\left( 2\Pi^{e\gamma^-}(x,z)+\Pi^{\gamma^-\gamma^-}(x,z)\right)=0\,. \label{o34}
\gal
\end{subequations}
for any $z$. It follows from \eq{o30},\eq{o33} and \eq{o34} after a simple calculation using \eq{n13} that 
\begin{subequations}\label{o38A}
\bal
C_e(z)&= 2-\frac{z^2}{2(z+\nu)^2}+ 2\ln\left( 1+\frac{\nu}{z}\right)\,,\label{o38}\\
C_\gamma(z) &= -\frac{1}{3}\sqrt{1-\frac{4\nu}{z}}\left(2+\frac{4\nu}{z}\right)\,.\label{o39}
\gal
\end{subequations}

In the chiral limit $\nu= 0$: $C_e=3/2$,  $C_\gamma = -2/3$.

Eqs.~\eq{o20A} with \eq{o6A},\eq{o38A} constitute the evolution equations of the chiral cascade in QED at any $m$.

%%%%%%%%%%%%%%%%%%%%%%%%%%%%%%
%%%%%%%%%%%%%%%%%%%%%%%%%%%%%%
\section{Chiral Cherenkov cascade in $\chi$QCD}\label{sec:QCD}

We do not need to repeat all the arguments that lead us to the evolution equations in $\chi$QED. Rather, we only need to list the appropriate splitting functions in the chiral limit. The corresponding amplitudes were computed in \cite{Hansen:2024rlj}. It should be noted that in addition to the conventional QCD vertices, the non-Abelian Chern-Simons term supplies an additional vertex proportional to $b_0$. However, in the high energy limit its contribution to the scattering amplitudes is suppressed by the powers of energy.

The leading contributions to the production rates  mediated by the chiral anomaly in QCD, in the high-energy limit, are given by:  
\begin{subequations}\label{QCD}
\bal
\frac{dW^{q\to g^+ q}}{dx}&= \frac{C_F}{\tau}\frac{1+(1-x)^2}{x}\,,\label{QCD1}\\
\frac{dW^{q\to q g^+}}{dx}&= \frac{C_F}{\tau}\frac{1+x^2}{1-x}\,,\label{QCD1a}\\
\frac{dW^{g^-\to q\bar q}}{dx}&= \frac{1}{2\tau}\left[x^2+(1-x)^2\right]\,,\label{QCD1b}\\
\frac{dW^{g^+\to g^+ g^+}}{dx}&= \frac{2N_c}{\tau}\frac{1}{x(1-x)}\,,\label{QCD1e}\\
\frac{dW^{g^-\to g^+ g^-}}{dx}&= \frac{2N_c}{\tau}\frac{(1-x)^3}{x}\,,\label{QCD1d}\\
\frac{dW^{g^-\to g^- g^+}}{dx}&= \frac{2N_c}{\tau}\frac{x^3}{(1-x)}\,,\label{QCD1c}
\gal
\end{subequations}
where $C_F= (N_c^2-1)/(2N_c)=4/3$. All other channels are suppressed by the inverse powers of energy \cite{Hansen:2024rlj}. The total $g\to gg$ rate reads\footnote{A useful formula for computing the rates for a given circular photon polarization is derived in Appendix~\ref{App-2}. }
\bal\label{r6}
\frac{dW^{g\to g g}}{dx}&= \frac{1}{2}\sum_{\lambda_1=\pm}\sum_{\lambda_2=\pm}\sum_{\lambda_3=\pm}
\frac{dW^{g^{\lambda_1}\to g^{\lambda_2} g^{\lambda_3}}}{dx}\,,
\gal
where $1/2$ is the average over polarizations of the initial gluon. The only non-vanishing partial rates are \eq{QCD1c}-\eq{QCD1e}. We define the real contributions (those with $x<1$) to the polarized splitting functions as
\bal\label{r8}
\frac{dW^{g\to g g}}{dx}= \frac{1}{\tau}\tilde P_{gg}(x)\,,\quad \frac{1}{2}\frac{dW^{g^{\lambda_1}\to g^{\lambda_2} g^{\lambda_3}}}{dx}= \frac{1}{\tau}\tilde P_{g^{\lambda_2}g^{\lambda_1}}(x)\,.
\gal
Eqs.~\eq{r6},\eq{r8} imply that the total $g\to g$ splitting function can be computed as 
\bal\label{r10}
P_{gg}(x)=\sum_{\lambda_1=\pm}\sum_{\lambda_2=\pm} P_{g^{\lambda_2}g^{\lambda_1}}(x)\,.
\gal
 Using \eq{QCD} and the definition \eq{r8} we find
\begin{subequations}\label{QCD2}
\bal
P_{g^+ q}(x)&= C_F\frac{1+(1-x)^2}{x}\,,\label{QCD2a}\\
P_{q q}(x)&=C_F\left[ \frac{1+x^2}{(1-x)_+}+\frac{3}{2}\delta(1-x)\right]\,,\label{QCD2c}\\
P_{q g^-}(x)&= \frac{1}{2}\left[x^2+(1-x)^2\right]\,,\label{QCD2b}\\
P_{g^+g^+}&=\frac{N_c}{x}\left[\frac{1}{(1-x)_+} +C^{+}\delta(1-x)\right]\,,\label{QCD2d}\\
P_{g^+g^-}&=\frac{N_c}{x}(1-x)^3\,,\label{QCD2e}\\
P_{g^-g^-}&=N_c\left[\frac{x^3}{(1-x)_+} +C^{-}\delta(1-x)\right]\,,\label{QCD2f}
\gal
\end{subequations}
and other splitting functions vanish. The evolution equations of the chiral Cherenkov cascade in QCD read:
\begin{subequations}\label{R20}
\bal
\frac{dq(y,t)}{dt}=&\frac{1}{\tau} \int_y^1\frac{dx}{x}\left\{ P_{qq}(x)q\left( \frac{y}{x},t\right)+P_{qg^-}(x)g^-\left( \frac{y}{x},t\right)\right\}\,,\label{r20}\\
\frac{d\bar q(y,t)}{dt}=&\frac{1}{\tau} \int_y^1\frac{dx}{x}\left\{ P_{qq}(x)\bar q\left( \frac{y}{x},t\right)+P_{qg^-}(x)g^-\left( \frac{y}{x},t\right)\right\}\,,\label{r21}\\
\frac{dg^+(y,t)}{dt}=&\frac{1}{\tau} \int_y^1\frac{dx}{x}\left\{N_f P_{g^+ q}(x)q\left( \frac{y}{x},t\right)+N_fP_{g^+ q}(x)\bar q\left( \frac{y}{x},t\right)\right.\nonumber\\
&+\left. 
P_{g^+g^+}(x)g^+\left( \frac{y}{x},t\right)
+P_{g^+g^-}(x)g^-\left( \frac{y}{x},t\right)
\right\}\,,\label{r22}\\
\frac{dg^-(y,t)}{dt}=&\frac{1}{\tau} \int_y^1\frac{dx}{x} P_{g^- g^-}(x)g^-\left( \frac{y}{x},t\right)\,,\label{r23}
\gal
\end{subequations}
where $q(x,t)$ and $\bar q(x,t)$ are quark and antiquark distributions for a single flavor. In the chiral limit all flavors have equal distributions.

To determine the virtual contributions at $x=1$, represented by the constants $C^+$ and $C^-$, we employ the charge and energy conservation as in the previous section. The charge conservation implies $A_1^{qq}=0$ and is satisfied by \eq{QCD2c}. The total energy of the system is the sum of the second moments of all distributions, see \eq{n38}. Its  conservation requires that 
\bal\label{r25}
\frac{d}{dt}\left[N_f( q_2 + \bar q_2)+  g^+_2 + g^-_2\right]=0\,,
\gal
which implies  that 
\begin{subequations}\label{R26}
\bal
A_2^{qq}+A_2^{g^+q}=0\,,\label{r26}\\
2N_fA_2^{qg^-}+A_2^{g^+g^-}+A_2^{g^-g^-}=0\,,\label{r27}\\
A_2^{g^+g^+}=0\,.\label{r28}
\gal
\end{subequations}
As in $\chi$QED \eq{r26} is satisfied automatically along with  the charge conservation. The other two equations \eq{r27},\eq{r28} yield 
\bal\label{r30}
C^+=0\,,\quad  N_cC^-= \frac{11N_c}{6}-\frac{N_f}{3}\,.
 \gal
It is not surprising that $C^+$ vanishes because the only decay channel for the right-handed gluons is $g^+\to g^++g^+$  and it clearly conserves their total energy.

Substituting \eq{QCD2d}-\eq{QCD2f} into \eq{r10} we obtain the familiar unpolarized gluon-gluon splitting function: 
\bal\label{r32}
P_{gg}(x)=2N_c\left[ \frac{1-x}{x}+\frac{x}{(1-x)_+}+x(1-x)\right] + \left( \frac{11}{6}N_c-\frac{N_f}{3}\right)\delta(1-x)\,.
\gal
The deviations from the standard QCD splitting functions appear when the mass terms are taken into account as in \eq{o6A}. However, we do not consider them in the present work.

\begin{figure}[ht]
\begin{tabular}{cc}
      \includegraphics[height=5.cm]{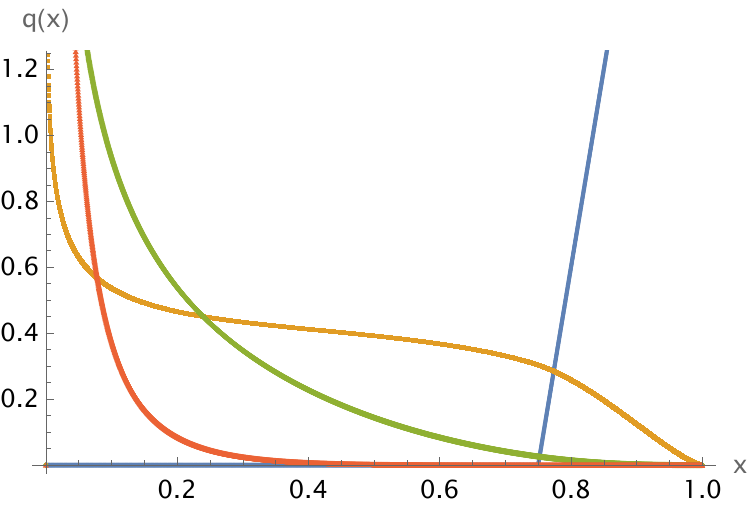} &
      \includegraphics[height=5.cm]{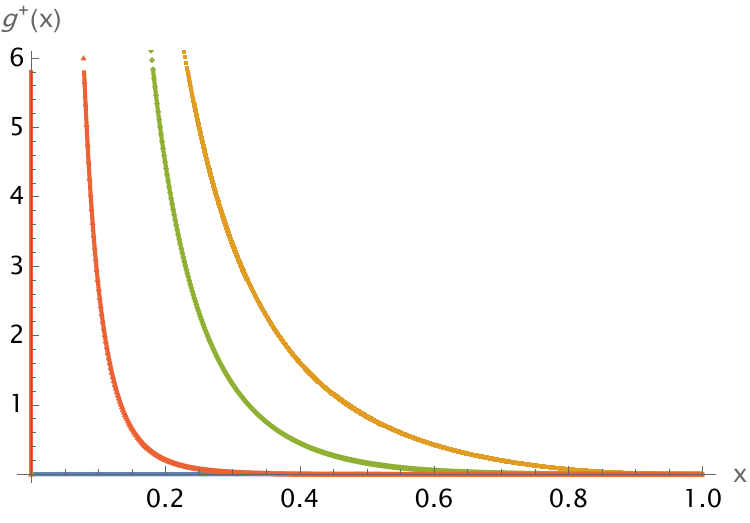}
      \end{tabular}
  \caption{Chiral Cherenkov cascade in $\chi$QCD with the initial conditions: $q(x,0)= 32 \left(x-3/4\right)H\left(x\ge 3/4\right)$, $\bar q(x,0)=g^\pm (x,0)=0$. Blue line: $t=0$, yellow line: $t=\tau/2$, green line: $t=\tau$, red line: $t=2\tau$.}
\label{fig:quark}
\end{figure}

\begin{figure}[ht]
\begin{tabular}{cc}
      \includegraphics[width=0.45\linewidth]{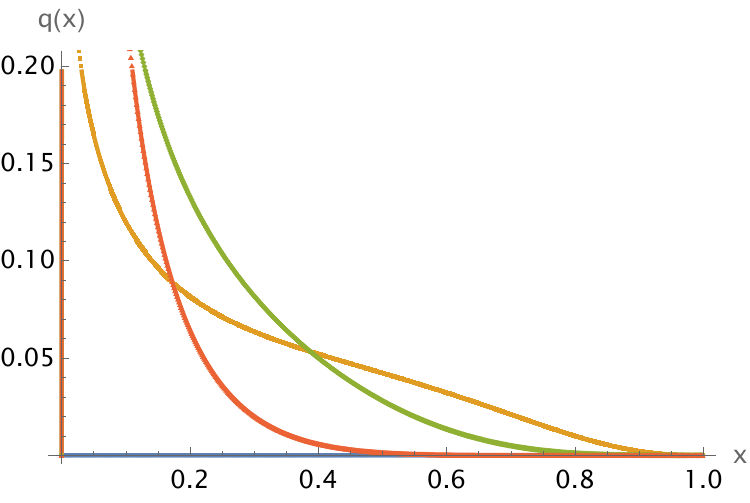} &
      \includegraphics[width=0.45\linewidth]{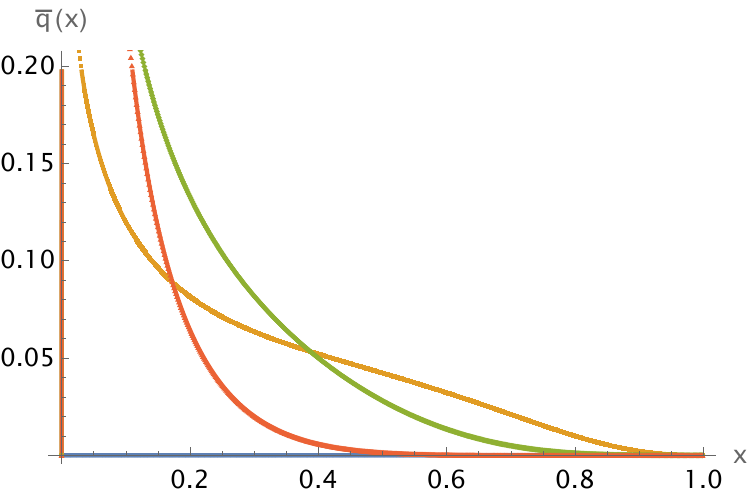}\\
      \includegraphics[width=0.45\linewidth]{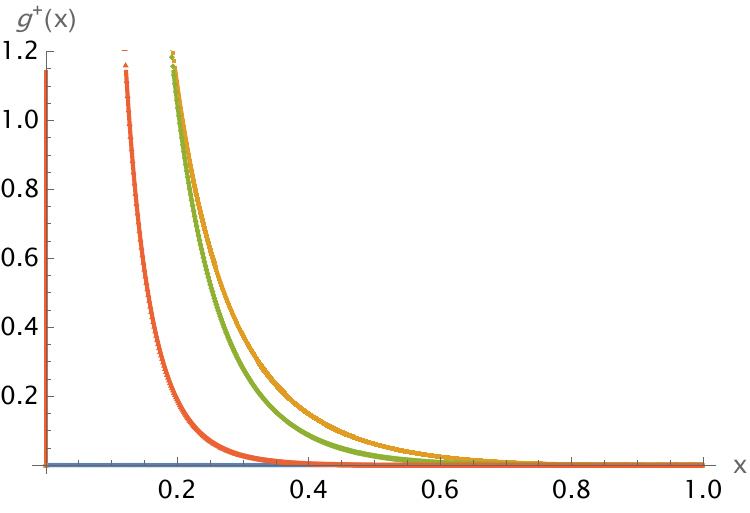} &
      \includegraphics[width=0.45\linewidth]{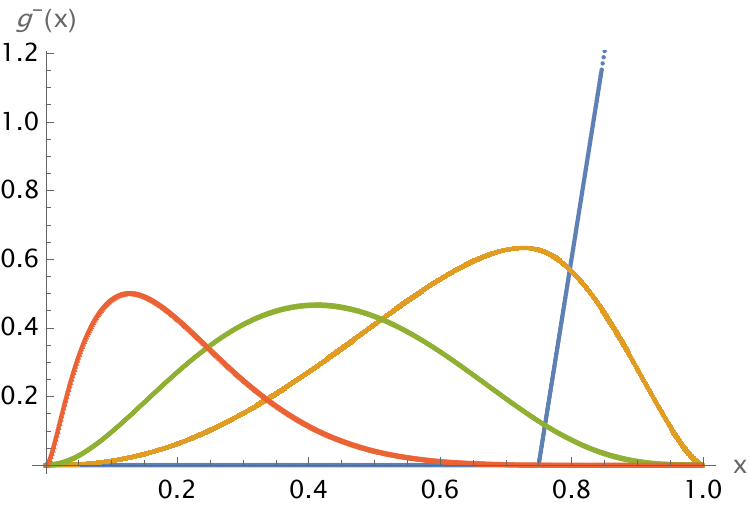}
      \end{tabular}
  \caption{Chiral Cherenkov cascade in $\chi$QCD with the initial conditions: $g^- (x,0)= 32 \left(x-3/4\right)\step\left(x\ge 3/4\right)$, $q(x,0)=\bar q(x,0)=g^+(x,0)=0$. Blue line: $t=0$, yellow line: $t=\tau/2$, green line: $t=\tau$, red line: $t=2\tau$. }
\label{fig:neggluon}
\end{figure}

\begin{figure}[ht]
\begin{tabular}{cc}
      \includegraphics[width=0.45\linewidth]{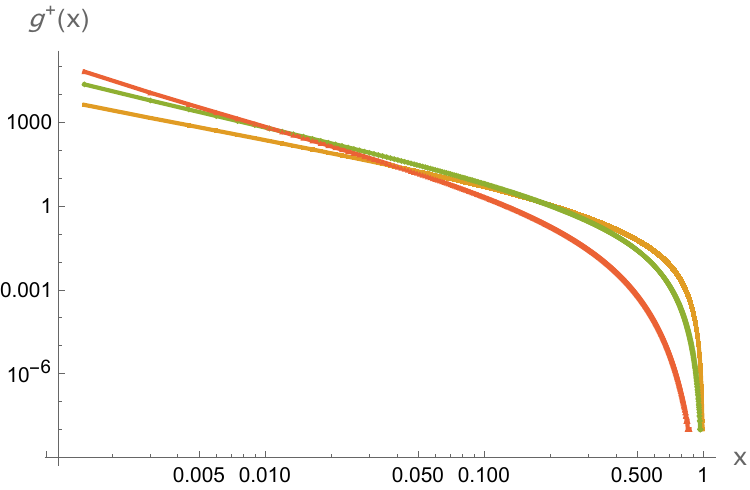} &
      \includegraphics[width=0.45\linewidth]{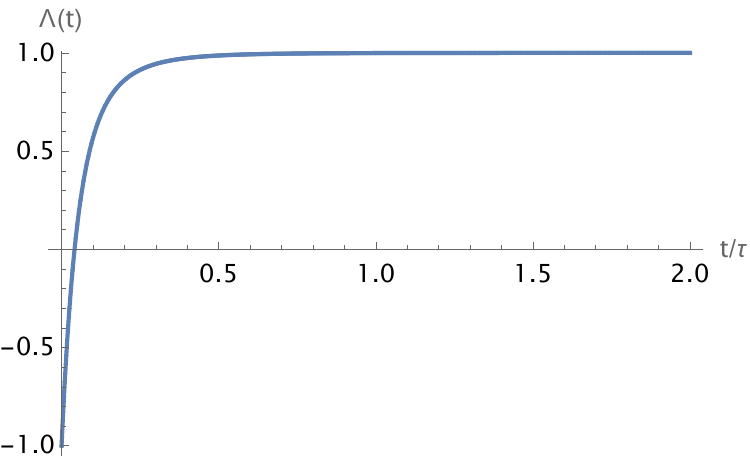}
      \end{tabular}
  \caption{Left panel: log plot for $g^+$. Right panel: The degree of polarization $\Lambda(t)$  of the cascade shown in  \fig{fig:neggluon}. }
\label{fig:lambda_q}
\end{figure}

Figs.~\ref{fig:quark} and \ref{fig:neggluon} show the time-evolution of the distributions functions in $\chi$QCD with two different initial conditions similar to those of Figs.~\ref{fig:electron} and \ref{fig:neglight} in $\chi$QED. The main difference between them is in the time-evolution of gluon and photon distributions. The additional gluon decay channel facilitates 
production of the low-$x$ right and left-handed gluons. As a result, the number of right-handed gluons by far exceeds the number  of the right-handed photons at the same $t/\tau$. Also, the left-handed gluons are distributed in the entire interval $0<x<1$ whereas the left-handed photons are found only with $x\ge 3/4$ fixed by the initial condition.  As in the case of photons, the total number of the right-handed gluons increases, whereas the number of left-handed ones decreases. Thus, the degree of the cascade polarization $\Lambda$ rapidly approaches unity, which happens to gluons an order of magnitude earlier  than to photons as seen  in the right panel of \fig{fig:lambda_q}.

%%%%%%%%%%%%%%%%%%%%%%%%%%%%%%
%%%%%%%%%%%%%%%%%%%%%%%%%%%%%%
\section{Summary and Discussion}\label{sec:summary}

We studied the time-evolution of cascades using the Chern-Simons extensions of QED and QCD, which allows systematic incorporation of  the chiral magnetic current $\b j = b_0 \b B$ in the perturbation theory. Our main result is the time-evolution equations \eq{N20}, \eq{o20A}, \eq{R20} for the distribution functions. These equations omit the contribution of the conventional processes which has been extensively studied in the literature 
\cite{Arnold:2002zm,Baier:2000sb,Blaizot:2015lma,Mehtar-Tani:2022zwf,Mueller:1999pi,Schlichting:2020lef,PhysRevLett.105.195005,Bulanov:2013cga,Barata:2021wuf,Caron-Huot:2010qjx,Sievert:2019cwq}. In \cite{Hansen:2022nbs,Hansen:2023wzp} we analyzed in detail the contribution of the chiral anomaly to the Bethe-Heitler process and determined the conditions when such a contribution becomes essential. 

The interplay of the conventional bremsstrahlung and pair production with their chiral Cherenkov counterparts is complex because they arise from different kinematic regions. For example, in contrast to the conventional QED/QCD bremsstrahlung which is dominated by soft photon/gluon emission, the chiral Cherenkov emission is determined by the resonance in the intermediate fermion states \cite{Hansen:2023wzp}. Unlike the former process that generates large logarithms of transverse momentum, the latter one generates large powers of the resonance relaxation time. The fact that the  conventional and the chiral Cherenkov processes come from different parts of the phase space makes it challenging to combine them in a single framework. One possible approach could involve the kinetic equation with the conventional and chiral Cherenkov processes contributing to different collision integrals.

The evolution equations \eq{N20}, \eq{R20} closely resemble the Gribov-Lipatov equations of QED \cite{Lipatov:1974qm} and the DGLAP equations of QCD \cite{Gribov:1972ri,Dokshitzer:1977sg,Altarelli:1977zs} in the leading logarithmic approximation. This is because the splitting functions are exactly the same in the high energy approximation and in the chiral limit. Nevertheless, they are fundamental different as the evolution equations of the chiral Cherenkov cascade \eq{N20}, \eq{R20} involve the two photon and gluon polarizations in highly asymmetric way. As a result the cascade quickly becomes strongly polarized in violation of parity.

In \cite{Hansen:2024rlj} we computed the rate of energy loss $-dE/dz$ by a fast parton in the quark-gluon plasma due to the chiral Cherenkov radiation of a single gluon. We found that for a reasonable value of $b_0$, this mechanism can compete with the conventional QCD process. Moreover, we observed that its energy dependence is similar to that of jet quenching in the Bethe-Heitler regime. Also, quantum interference effects are known to suppress the energy dependence of the conventional mechanism, whereas the chiral processes we considered in this paper are already coherent throughout the plasma volume. Thus, we anticipate that the relative contribution of the anomalous processes to the energy loss will increase with energy. An accurate estimation of the impact of the chiral anomaly on jet quenching necessitates the incorporation of the chiral Cherenkov processes into the phenomenological frameworks.

%%%%%%%%%%%%%%%%%%%%%%%%%%%%%%
%%%%%%%%%%%%%%%%%%%%%%%%%%%%%%
\begin{acknowledgments}
This work  was supported in part by the U.S. Department of Energy Grants No.\ DE-SC0023692.
\end{acknowledgments}

%%%%%%%%%%%%%%%%%%%%%%%%%%%%%%
%%%%%%%%%%%%%%%%%%%%%%%%%%%%%%
\section*{Data Availability}

The data that support the findings of this article are not
publicly available. The data are available from the authors
upon reasonable request.

%%%%%%%%%%%%%%%%%%%%%%%%%%%%%%
%%%%%%%%%%%%%%%%%%%%%%%%%%%%%%
\bibliography{anom-biblio}

%%%%%%%%%%%%%%%%%%%%%%%%%%%%%%
%%%%%%%%%%%%%%%%%%%%%%%%%%%%%%
%\newpage
\appendix

\section{Double photon emission}\label{Adpe}

In \sec{sec:dbl} we argued that the intermediate states in the multiple-emission processes can be put on mass-shell at large relaxation times $\tau$. In this Appendix we develop a more detailed argument considering as an example  the double photon emission $e\to e\gamma\gamma$. We argue that the radiation rate $W^{e\to e\gamma\gamma}$ is dominated by a single pole induced by the chiral anomaly. 

Let $p$, $p'$, $k_1$, $k_2$ be the 4-momenta of the initial and final electrons and the two photons respectively. At high energy they read:  
\begin{subequations}\label{C1a}
\bal
&p=\left(E,0,0,E-\displaystyle \sum_{i=0}^{1}\frac{k_{i\perp}^2+x_i(1-x_i)\mu^2(E)}{2x_i(1-x_i)E}\right)\,,\label{C1aa}\\
&k_i=\left(z_i E,k_{i\perp},0,x_i E-\frac{k_{i\perp}^2+\mu^2(x_i)}{2x_i E}\right)\,,\label{C1ab}\\
&p'=\left(E',-k_{0\perp}-k_{1\perp},0,E'-\displaystyle \sum_{i=0}^{1}\frac{k_{i\perp}^2+\mu^2(E')}{2E'}\right) \,.\label{C1ac}
\gal
\end{subequations}
The matrix element can be written as
\bal
i\mathcal{M}^{e\to e\gamma\gamma}&\approx- g\bar u_{\b p' s'}\slashed{e}^*_{\b k_1 \lambda_1}u_{\b p' s'}\frac{i}{(p'+k_1)^2-m^2} \bar u_{\b p s}\slashed{e}^*_{\b k_0 \lambda_0} u_{\b p s}\nonumber\\&=ig\bar u_{\b p' s'}\slashed{e}^*_{\b k_1 \lambda_1}u_{\b p' s'}\frac{i}{(p'+k_1)^2-m^2} i\mathcal{M}^{e\to e\gamma}\,.\label{DD1}
\gal
Using explicit expressions for $ u_{\b p' s'}$ and $\slashed{e}^*_{\b k_1 \lambda_1}$ we can obtain an expression for the square of the matrix element
\bal
|\mathcal{M}^{e\to e\gamma\gamma}|^2&\approx g^2\frac{2(1-x_1)\b k_{1\perp}^2}{( k_{1\perp}^2+\mu^2(x_1)(1-x_1)+m^2x_1^2)^2}\left[\frac{1+(1-x_1)^2}{x_1}+\frac{m^2x_1^3}{\b k_{1\perp}^2}\right]|\mathcal{M}^{e\to e\gamma}|^2\,.\label{DD2}
\gal
The rate is therefore
\bal
W^{e\to e\gamma\gamma}(y)\approx 2g^2\int \frac{dx_1d\b k_{1\perp}^2}{16\pi^2}\frac{\b k_{1\perp}^2}{( k_{1\perp}^2+\mu^2(x_1)(1-x_1)+m^2x_1^2)^2}\nonumber\\
\times\left[\frac{1+(1-x_1)^2}{x_1}+\frac{m^2x_1^3}{k_{1\perp}^2}\right]dW^{e\to e\gamma}\,,\label{DD3}
\gal
where $x_1$ is bounded from below  $y$. The divergence is regulated by the relaxation time $\tau$. The contribution from the pole becomes 
\bal
W^{e\to e\gamma\gamma}_\text{pole}(y)\approx \frac{g^2\tau}{4\pi E}\int \frac{dx_1\b k_{1\perp}^2d\b k_{1\perp}^2}{x_1 (1-x_1)}\delta( k_{1\perp}^2+\mu^2(x_1)(1-x_1)+m^2x_1^2)\nonumber\\
\times\left[\frac{1+(1-x_1)^2}{x}+\frac{m^2x_1^3}{k_{1\perp}^2}\right]dW^{e\to e\gamma}\,.\label{DD4}
\gal
Integrating over $k_{1\perp}$ yields 
\bal
W_{e\to e\gamma\gamma}^\text{pole}(y)\approx \frac{g^2\tau}{4\pi E}\int_y^{x_0} \frac{dx_1}{x_1^2}
\left[-\mu^2(x_1)(1+(1-x_1)^2)-2m^2x_1^2\right]dW_{e\to e\gamma}\,.\label{DD5}
\gal
At this point, one needs to be careful when integrating over $x_1$ due to the fact that $dW^{e\to e\gamma}$ may depend on it. Luckily the parameter $b_0$ is believed to be small in comparison to the fermion's mass. If $b_0$ is small enough such that $b_0 E\ll m^2$. In this limit, $x^q_+\ll 1$, and $W^{e\to e\gamma}$ can be pulled out of the integral. After integrating over frequencies for $\lambda_1=1$ and $\mu^2(z_1)=-b_0 E x_1$ gives
\bal
W^{e\to e\gamma\gamma}_\text{pole}(y)&\approx \frac{2g^2W_{e\to e\gamma} \tau b_0 }{4\pi }   \left[\log{\frac{x_0}{y}}+\frac{(y-3x_0)(x_0-y)}{4x_0}\right]\,.\label{DD6}
\gal
 In the soft photon limit such that $y\ll x_0$ we have:
\bal
W^{e\to e\gamma\gamma}_\text{pole}(y)&\approx  \frac{2g^2W_{e\to e\gamma} \tau b_0 }{4\pi }   \log{\frac{x_0}{y}}\,,\label{DD7}
\gal
which can be compared to the contribution away from the pole.

Now consider the behavior away from the pole such that $ k_{1\perp}^2+\mu^2(x_1)(1-x_1)+m^2z_1^2>\frac{E}{\tau}$. In this region \eq{DD4} needs no regulator, and the integrals may be taken directly. Integration over $\b k_\perp^2$,
\bal
W^{e\to e\gamma\gamma}_\text{away}(y)&\approx \frac{g^2W_{q\to qg}}{8\pi^2}\int_y^{x_0} \frac{dx_1}{x_1}(1+(1-x_1)^2)\log{\frac{K_\bot^2\tau}{E}}\,,\label{De1}
\gal
where $K_\bot^2$ is the upper bound of $k_{1\perp}^2$. Integrating over frequencies
\bal
W^{e\to e\gamma\gamma}_\text{away}(y)&\approx \frac{g^2W^{e\to e\gamma}\step\left(E\ge E_1\right)}{3\pi^2} \left[\log{\frac{x_0}{y}}+\frac{x_0^2-y^2}{2}-4(x_0-y)\right]\log{\frac{K_\bot^2\tau}{E}}\,.\label{De2}
\gal
For soft photons this simplifies to 
\bal
W^{e\to e\gamma\gamma}_\text{away}&\approx \frac{g^2W^{e\to e\gamma}\step\left(E\ge E_1\right)}{3\pi^2} \log{\frac{K_\bot^2\tau}{E}}\log{\frac{x_0}{y}}\,.\label{De3}
\gal
This can be compared to the contribution from the pole:
\bal
\frac{W^{e\to e\gamma\gamma}_\text{away}}{W^{e\to e\gamma\gamma}_\text{pole}}&\approx \frac{\step\left(E\ge E_1\right)}{2\pi b_0 \tau} \log{\frac{K_\bot^2\tau}{E}}\,.\label{De}
\gal
Therefore if $\tau\gg \frac{1}{2\pi b_0 } \log{\frac{K_\bot^2\tau}{E}}$ then the pole gives the dominate contribution. For $\frac{1}{\tau}=\frac{\alpha b_0}{4}$ this implies that $\frac{8 \pi}{\alpha}\gg \log{\frac{4 K_\bot^2 }{\alpha b_0 E}}$ which holds for nearly any $E$.

This procedure is not restricted to the double photon production. For any given process the anomaly introduces a new pole for every intermediate particle involved. As such, each of these particles can be approximated to be on mass-shell. The justification in each case is very similar. The contributions to their corresponding radiation rates are dominated by such poles.

%%%%%%%%%%%%%%%%
\section{Circular polarization tensor}\label{App-2}

Define the photon polarization sum as:
\bal\label{t1}
d^{\mu\nu}_\lambda=(\epsilon_{\b k,\lambda}^{\mu})^*\epsilon_{\b k,\lambda}^\nu\,, \quad \text{(no summation)}\,.
\gal
The polarization vectors describe the right and left-hand circular polarizations and obey the identity $i\b k\times \b\epsilon_{\b k,\lambda}= \lambda |\b k|\b\epsilon_{\b k,\lambda}$. In the radiation gauge the polarization vectors satisfy $\epsilon_{\b k,\lambda}^0=0$,  
$\b\epsilon_{\b k,\lambda}\cdot \b k =0$,  $\b (\epsilon_{\b k,\lambda})^*\cdot \b \epsilon_{\b k,\lambda'}= \delta_{\lambda,\lambda'}$.  It follows that $d^{00}_\lambda= d^{0i}_\lambda=0$, $d^{ij}_\lambda k^i=0$, $\Tr \sum_\lambda d^{ij}_\lambda=2$ and 
$d^{ij}_\lambda \epsilon_{\b k,\lambda}^i= \epsilon_{\b k,\lambda}^j$ (no summation over $\lambda$). The space-space components of \eq{t1} that satisfy these constraints read:
\bal\label{t3}
d^{ij}_\lambda=(\epsilon_{\b k,\lambda}^{i})^*\epsilon_{\b k,\lambda}^j
=\frac{1}{2}\left( \delta^{ij}-\frac{k^ik^j}{\b k^2}\right) +\frac{i\lambda}{2}\varepsilon^{ij\ell}\frac{k^\ell}{|\b k|}\,.
\gal 
See also Appendix in \cite{Tuchin:2020pbg}.

In the collinear limit one can verify \eq{t3}  by using the explicit form of the polarization vector $\epsilon_{\b k, \lambda}= (0,\b \epsilon_\bot,\epsilon_z)$ with $\b \epsilon_\bot= (\unit x+i\lambda \unit y)/\sqrt{2}$, $\epsilon_z= -\b \epsilon_\bot\cdot \b k/k_z$, correct up to small $k_\bot^2/k_z^2$ terms:
\bal\label{t5}
d^{ij}_\lambda=\left( \frac{1}{\sqrt{2}}, \frac{-i\lambda}{\sqrt{2}},\epsilon_z^*\right)^T
\left( \frac{1}{\sqrt{2}}, \frac{i\lambda}{\sqrt{2}},\epsilon_z\right)=
\frac{1}{2}\left( 
\begin{array}{ccc}
1 & i\lambda & \frac{-k_x-i\lambda k_y}{k_z}\\
-i\lambda & 1 & \frac{-k_y+i\lambda k_x}{k_z}\\
\frac{-k_x+i\lambda k_y}{k_z} & \frac{-k_y-i\lambda k_x}{k_z} & 0
\end{array}
\right) + \mathcal{O}\left( \frac{k_\bot^2}{k_z^2}\right)
\gal
which agrees with \eq{t3} in the same limit. 

\end{document}